\documentclass[nofootinbib,showkeys,preprint,aps]{revtex4-1}

\usepackage{amsmath,amsfonts}
\usepackage{array}
\usepackage{braket} %please add this package we need it.
\usepackage{hyperref}
\usepackage{color}  % for color

\usepackage{amssymb}
\begin{document}

\title{ The Ginzburg-Landau Theory of a Holographic Superconductor}

\author{Lei Yin}
\affiliation{Institute of Particle Physics, Huazhong Normal University,Wuhan 430079, China}
\email{lei@iopp.ccnu.edu.cn}

\author{Defu Hou}
\affiliation{Institute of Particle Physics, Huazhong Normal University, Wuhan 430079, China}
\email{defu@iopp.ccnu.edu.cn}

\author{Hai-cang Ren}
\affiliation{Physics Department, The Rockefeller University,1230 York Avenue, New York, 10021-6399, U.S.A.}
\email{ren@mail.rockefeller.edu}

\begin{abstract}
The general Ginzburg-Landau formulation of a holographic superconductor is developed  near the transition temperature in the probe limit for two kinds of conformal dimension. Below the transition temperature, $T<T_c$, the order-parameter scales with $\sqrt{1-\frac{T}{T_c}}$ as expected. The analytical expressions of  G-L free energy in grand canonical ensemble and canonical ensemble are derived and the gradient term is studied. Furthermore this scaling coefficient of order-parameter takes different values in the grand canonical ensemble and the canonical ensemble, suggesting the strong coupling nature of the boundary field theory of the superconductivity.
\end{abstract}
%\pacs{111}


\maketitle

\section{Introduction}
\label{sec-1}

The holographic principle, proposed by 't Hooft and Susskind, \cite{hooft_dimensional_1993, susskind_world_1994} has been widely recognized as a promising probe of some
universal properties of strongly coupled systems. It is conjectured that a strongly coupled quantum system can be regarded as the holographic image of a weakly coupled gravitation theory in higher spatial dimensions and thereby becomes
analytically tractable. One example of the holographic principle is the correspondence
between the supergravity in AdS$_5$ background and the $\mathcal{N}=4$ super Yang-Mills theory in $M_4$ in the limit
of large $N_c$(the number of colors) and large 't Hooft coupling \cite{maldacena_large_1998,wittern_anti-de_1998,gubser_gauge_1998,klebanov_ads/cft_1999,aharony_large<_2000}. Among the success of the application of this
correspondence to the quark-gluon plasma produced in the relativistic heavy ion collisions (RHIC) are the equation  of state, the viscosity ratio \cite{policastro_shear_2001}.

Another playground of the holographic principle covers the strongly correlated condensed matter systems. The holographic
superconductivity (HSC) proposed in Refs.   \cite{gubser_breaking_2008,gubser_gravity_2008,hartnoll_holographic_2008,horowitz_holographic_2008, hartnoll_building_2008, herzog} is an interesting attempt in this direction. To simulate the layer structure of some high temperature superconductors such as cuprates, the thermodynamics
in 2+1 dimensions (M$_3$) is modeled as the boundary physics of the gravity in an AdS$_4$ (with the M$_3$ as the AdS-boundary) black hole background coupled to a $U(1)$
gauge field and a complex scalar field. When the Hawking temperature $T$ is lowered below a critical value, $T_c$,
the scalar field acquires a nontrivial configuration in the bulk whose image at the AdS-boundary corresponds to the long range order parameter
of the superconductivity. An important result of this model is that the real part of the AC conductivity of HSC at $T=0$ shows a gap in frequency \cite{horowitz_holographic_2008} for
$0 < \omega < \omega_0$ with
\begin{equation}
\omega_0/T_c\approx 8,
\label{AC}
\end{equation}
in remarkable agreement with the experimental value of cuprate
superconductors. This larger ratio compared to the BCS prediction, $\omega_0/T_c\approx 3.5$, suggests the strong coupling nature
of the holographic superconductor.
Up to now, extensive investigations have been carried out for HSC in the literature. The original HSC model has been
generalized to include non s-wave order parameters at the boundary and to higher powers of the curvature, the higher order Maxwell field and diverse kinds of black holes in the bulk \cite{maeda_characteristic_2008,maeda_universality_2009,gregory_holographic_2009,horowitz_zero_2009,barclay_gauss-bonnet_2010,setare_holographic_2010,cai_holographic_2010,ge_analytical_2010,benini_gauge_2010,chen_towards_2010,jing_holographic_2010,zeng_d-wave_2010,zeng_analytical_2010,chen_analytic_2011,kanno_note_2011,jing_holographic_2011,ge_analytical_2011,pan_analytical_2012,gangopadhyay_analytic_2012-2,gangopadhyay_analytic_2012,momeni_holographic_2013,yao_analytical_2013,li_gauss-bonnet_2013,herzog_analytic_2010}.

In this paper, we derive the  Ginzburg-Landau (GL) free energy of HSC in the probe limit. We obtain analytic
expressions of the GL coefficients in both grand canonical ensemble and canonical ensemble with conformal dimension one and two, respectively.   As the critical temperature is approached from below, we find the order parameter
\begin{equation}
\Braket{\mathcal{O}_\triangle} \approx \kappa\ T_c^\triangle\sqrt{1-\frac{T}{T_c}}.
\label{eq:GL_order}
\end{equation}
along with an analytical formula of the coefficient $\kappa$, where $\triangle(=1,2)$ denotes the
conformal dimension and the dependence of $\kappa$ and $T_c$ on $\triangle$ is not indicated
explicitly. Numerically, we have
\begin{equation}
\begin{aligned}
 T_c &= 0.214\ \mu,  &\kappa = 16.37 \quad &\text{in grand canonical ensemble} \\
 T_c &= 0.226\ \sqrt{\rho},  &\kappa = 9.451 \quad  &\text{in canonical ensemble}
\end{aligned}
\end{equation}
for $\triangle=1$ and
\begin{equation}
  \begin{aligned}
    T_c &= 0.0587\ \mu, &\kappa = 163.68 \quad &\text{in the grand canonical ensemble}\\
    T_c &= 0.118\ \sqrt{\rho}, &\kappa = 143.574 \quad  &\text{in the canonical ensemble}
\end{aligned}
\end{equation}
for $\triangle=2$, where $\mu$ and $\rho$ are the chemical potential and the density respectively.
While the scaling behavior $\sqrt{1-\frac{T}{T_c}}$, the numerical values of the coefficient and the critical temperature $T_c$ in canonical ensemble reported  in the literature
\cite{hartnoll_building_2008, hartnoll_holographic_2008} are reproduced. Our formula for the coefficient $\kappa$ disagree with that derived in the Ref.
\cite{siopsis_analytic_2010}. So our work is not a mere reformulation of existing knowledge on the subject.
Furthermore, in contrast to the GL theory of a BCS superconductor, where $\kappa\simeq 3.0633$ for
$\Braket{\mathcal{O}_1}\equiv\Delta$ or $\Braket{\mathcal{O}_2} \equiv T_c\ \Delta$ with $\Delta$ the energy gap, in both grand canonical and canonical ensembles,
the difference between the $\kappa$ values in the two ensembles resonates with the strong coupling
information implied by the ratio (\ref{AC}). Beyond the probe limit, our approach to the GL theory can be readily generalized to include the back-reaction of the scalar field to the gravity.

This paper is organized as follows. The general formulation and the probe limit are reviewed in the next section, where we
shall introduce the grand canonical ensemble and the canonical ensemble, and the order-parameter on two kinds of conformal dimension. The GL free energy in the grand canonical ensemble is
derived in the section 3 and that in the canonical ensemble in the section 4, both for a homogeneous condensate. The gradient term with an inhomogenous order parameter will be considered in the section 5. In the final section, we shall summarize and discuss our results,
where we shall pinpoint the reason for the difference between our coefficients in (\ref{eq:GL_order}) and those in the literature, and further justify our values.
The GL of a BCS superconductivity is presented in the appendix to illustrate the difference between the weak coupling and the strong coupling.

\section{The General Formulation}
\label{sec-2}
\subsection{The action and the equations of motion}
\label{sec-2-1}

The gravity dual in 3+1 dimensions of a holographic superconductor(HSC) in 2+1 dimensions consists of a metric field $g_{ab}$ with a negative cosmological constant $\Lambda = - \frac{6}{L^2}$;  a charged complex scalar field $\Psi$
with mass $m$ and charge $q$, and a $U(1)$ gauge potential $A_\mu$. The metric
$g_{ab}$ is asymptotically AdS$_4$. The classical action reads:
\begin{equation}
S_{_\text{HSC}} = S_{_\text{grav.}} + S_{_\text{matter}}+ S_{_\text{b}}
\label{eq:action-HSC}
\end{equation}
where
\begin{equation}
S_{_\text{grav.}} = \int \mathrm{d}^4x\sqrt{-g}\mathcal{L}_{_\text{grav.}}
=\frac{1}{16\pi G_N}\int \mathrm{d}^4x\ \sqrt{-g} \left( R - \frac{6}{L^2} \right)
\end{equation}
and
\begin{equation}
S_{_\text{matter}} = \int \mathrm{d}^4x\sqrt{-g}\mathcal{L}_{_\text{matter.}}
=\int \mathrm{d}^4x\ \sqrt{-g} \left( - \frac{1}{4}F_{\mu\nu}F^{\mu\nu} - |\nabla_\mu \Psi -i q A_\mu\Psi|^2 - m^2|\Psi|^2 \right) \label{eq:action}
\end{equation}
with $\mathcal{L}_{_\text{grav.}}$($\mathcal{L}_{_\text{matter.}}$) the Lagrangian density and $G_N$ the Newton's constant. The equations of motion follow from the variational principle. The derivatives of the variations of the field components,
$\delta g_{ab}$, $\delta A_\mu$ and $\delta\Psi$, in $\delta S_{_\text{grav.}}+\delta S_{_\text{matter}}$ can be transferred from the bulk to the boundary via integration by parts.
$S_{_\text{b}}$, residing at the boundary, is chosen such that its variation cancels these derivative terms. Besides the well-known Gibbons-Hawking term, which cancel the derivative of the metric at the boundary, $S_\text{b}$ may also contain the counter-terms involving the matter fields depending on the particular boundary conditions imposed and we shall come to this in due course \cite{henneaux2004asymptotically,henneaux2007asymptotic,witten2001multi} .  In this paper
we shall take $m^2 = - \frac{2}{L^2}$ to fit the Breitenlohner-Freedman condition. Before the section 5, we shall only consider the case where all matter fields depends only on $z$ with the 1-form $A_\mu \mathrm{d}x^\mu
= \Phi(z) \mathrm{d}t$
and $\Psi(z)$ a real-valued function.

The probe limit considered in this paper amounts to the limit that $q\to\infty$. Upon the scaling transformation
\begin{align}
\Psi = \frac{\tilde{\Psi}}{qL} ; \qquad  \Phi = \frac{\tilde{\Phi}}{q}
\end{align}
we have $S_\text{matter}=O\left(\dfrac{1}{q^2}\right)$ and the Einstein equation:
\begin{equation}
R_{ab}-\frac{1}{2}R g_{ab}-\frac{3}{L^2}g_{ab} = -8\pi G_NT_{ab} = O\left(\frac{1}{q^2}\right) \label{eq:Einstein_equation}
\end{equation}
for $\tilde{\Psi}=O(1)$ and $\tilde{\Phi}=O(1)$.
Thus we can split the metric tensor into two parts:
\begin{align}
g_{ab} = \bar{g}_{ab} + \delta g_{ab} \quad \text{and  } \quad \delta g_{ab} = O\left(\frac{1}{q^2}\right)
\end{align}
where $\bar{g}_{ab}$ is the solution of the vacuum Einstein equation, and $\delta g_{ab}$ denotes the matter field contribution. If the action (\ref{eq:action}) is
expanded in the powers of $\delta g_{ab}$, the 1st-order variation vanishes due to the
Einstein equation and then
\begin{align}
S_{_\text{HSC}} = \bar{S}_{_\text{HSC}} + O\big((\delta g_{ab})^2\big) = \bar{S}_{_\text{HSC}} + O\big( \frac{1}{q^4} \big)
\end{align}
where $\bar S_{_\text{HSC}}$ denotes $S_{_\text{HSC}}$ with $g_{ab}=\bar g_{ab}$.
Therefore in the probe limit $q \rightarrow +\infty$, the feedback of the matter field to the metric can be ignored and we are left with only
the vacuum Einstein-Hilbert action and the matter action plus some boundary terms to the order $O\left(\frac{1}{q^2}\right)$. With the metric independent of the  matter field, we need only to focus on $S_\text{matter}$ plus the related boundary terms with $\frac{1}{q^2}$ factored out. At this point $q$ may be set to one for convenience.

In the Poincaré coordinate system, $\{t,\ z,\ x_1,\ x_2\}$,  the metric tensor $\bar g_{ab}$ is that of the Schwarzschild-AdS$_4$ space-time and reads:
\begin{align}
 \mathrm{d}s^2 = \bar g_{ab}\ \mathrm{d}x^a \mathrm{d}x^b = \frac{L^2}{z^2}\bigg[f(z) \mathrm{d}t^2 + \frac{\mathrm{d} z^2}{f(z)}
+ \mathrm{d}x_1^2 + \mathrm{d}x_2^2\bigg]
\end{align}
where the metric function $f(z) = 1-\frac{z^3}{z_h^3}$ with $z_h$ the black hole horizon. The Hawking temperature of the black hole is $T = \frac{3}{4\pi z_h}$
and the AdS boundary is the 2+1 Minkowski space-time at $z=0$. We notice that in terms of $z_h$, $\tilde\Phi$ and $\tilde\Psi$, $S_{_\text{matter}}$ becomes $L$-independent, i.e.
\begin{equation}
\begin{aligned}
S_{_\text{matter}} &= \int \mathrm{d}t\int \mathrm{d}^2\vec x\int_0^{z_h} \mathrm{d}z\ \sqrt{-g} \mathcal{L}_\text{matter}[\Phi,\Psi]  \\
&= V \int \mathrm{d}t\ \int_0^{z_h}\mathrm{d}z \bigg( \frac{1}{2}\big(\frac{\mathrm{d}\Phi}{\mathrm{d}z}\big)^2
- \frac{f}{z^2} \big(\frac{\mathrm{d}\Psi}{\mathrm{d}z}\big)^2
+ \frac{\Psi^2\Phi^2}{z^2 f} + \frac{2\Psi^2}{z^4}\bigg)
\label{eq:action_L}
\end{aligned}
\end{equation}
where $V\equiv\int \mathrm{d}^2\vec x$ is the 2D spatial volume at the boundary and the tildes over $\Phi$ and $\Psi$ have been suppressed.
The arguments of $\mathcal{L}_{_\text{matter}}$ is indicated explicitly in (\ref{eq:action_L}) for later references. The classical equations of motion are:
\begin{align}
\frac{\delta S_{_\text{matter}}}{\delta \Phi} = 0 & \Rightarrow \;\; \frac{\mathrm{d}^2 \Phi}{\mathrm{d}z^2} = \frac{2\Psi^2}{z^2f} \Phi \label{eq:eom-phim} \\
\frac{\delta S_{_\text{matter}}}{\delta \Psi} = 0 & \Rightarrow \;\;\frac{\mathrm{d}}{\mathrm{d}z}\bigg[\frac{f}{z^2}\frac{\mathrm{d}  }{\mathrm{d}z} \Psi\bigg]
+ \frac{2\Psi}{z^4} + \frac{\Phi^2}{z^2f}\Psi = 0 \label{eq:eom-psim}
\end{align}
In order for the action to be finite, $\Phi$ has to vanish at the horizon $z=z_h$ and $\Psi$ has to be finite there.

The formulation may be further simplified with the horizon radius scaled to one via $z=z_h y$ and $\Phi=z_h^{-1}\tilde\Phi$. It follows that:
\begin{equation}
 S_{_\text{matter}} = V z_h^{-3}\int \mathrm{d}t\ \int_0^1\mathrm{d}y \bigg( \frac{1}{2}\big(\frac{\mathrm{d} \tilde\Phi}{\mathrm{d}y}\big)^2
- (\frac{1}{y^2} - y)\big(\frac{\mathrm{d}\Psi}{\mathrm{d}y}\big)^2
+ \frac{\Psi^2 \tilde\Phi^2}{y^2 - y^5} + \frac{2\Psi^2}{y^4}\bigg)
\label{eq:matter-action}
\end{equation}
with the horizon $y=1$.

\subsection{The thermodynamics}
\label{sec-2-2}

It follows from the holographic principle that the thermodynamics of a strongly correlated
field theory on 2+1 dimensions boundary is the image at the boundary of a weakly coupled gravity dual
in 3+1 dimensions bulk with the Euclidean signature, which is asymptotically AdS and holds a stationary black hole. In the case of the probe limit discussed in this paper, the gravity field decouples from the matters fields.
Thus, the thermodynamic partition function reads
\begin{align}
Z(T, \mu, V) = \text{const.}\int\mathcal{D}(\Phi_E) \mathcal{D}(\Psi)\; e^{-I[\Phi_E,\Psi]}
\approx e^{-\breve{I}}
\label{dual}
\end{align}
where $I$ is the Euclidean action of the gravity dual in the bulk and $T$ is the Hawking temperature of the black hole. $\breve{I}$ stands for the action $I$ evaluated at the saddle point, i.e. the solution of the equations of motion.
For the holographic superconductivity:
\begin{equation}
\begin{aligned}
I[\Phi_E, \Psi] & \equiv -\beta\int \mathrm{d}^2\vec x\int_0^{z_h} \mathrm{d}z\ \sqrt{g_{_E}}\ \mathcal{L}_{_\text{matter}}[-i\Phi_E,\Psi] + I_b[\Phi_E,\Psi] \\
               &= V\beta\int_0^{z_h} \mathrm{d}z  \bigg[\frac{1}{2}\big(\frac{\mathrm{d}\Phi_E}{\mathrm{d}z}\big)^2
                  + \frac{f}{z^2}\big(\frac{\mathrm{d}\Psi}{\mathrm{d}z}\big)^2 + \frac{\Psi^2\ \Phi_E^2}{z^2f} - \frac{2\Psi^2}{z^4} \bigg]
                  + I_b[\Phi_E,\Psi] \label{eq:euclidean}
\end{aligned}
\end{equation}
where a Wick rotation has been carried out to replace $\int \mathrm{d}t$ by $-i\beta$ with $\beta=\frac{1}{T}$ and $I_b$ is the Euclidean version of  $S_\text{b}$ necessary to produce the equations of motion
\begin{align}
\frac{\delta I}{\delta \Phi_E} = 0 \quad  \Rightarrow \quad  & \quad  \frac{\mathrm{d}^2\Phi_E}{\mathrm{d}z^2} = \frac{2\Psi^2}{z^2f} \Phi_E \label{eq:eom-phi} \\
\frac{\delta I}{\delta \Psi} = 0\quad  \Rightarrow \quad & \quad  \frac{\mathrm{d}  }{\mathrm{d}z}\bigg[\frac{f}{z^2}\frac{\mathrm{d}  }{\mathrm{d}z} \Psi\bigg] + \frac{2\Psi}{z^4} - \frac{\Phi_E^2}{z^2f}\Psi =0 \label{eq:eom-psi}
\end{align}

The saddle point dominating the path integral (\ref{dual}) corresponds to an imaginary $\Phi_E = i\Phi$
with real $\Phi$ and $\Psi$ satisfying (\ref{eq:eom-phim}) and (\ref{eq:eom-psim}).
In what follows, we shall refer to the solutions of equations of motion as being on-shell and to general
field configurations as being off-shell.

The chemical potential $\mu$ and the charge density $\rho$ can be extracted from the asymptotic behavior
\begin{align}
\Phi(z) = -i\Phi_E(z) \approx \mu - \rho z \qquad \text{as } z \rightarrow 0.
\label{asympt:phi}
\end{align}
As $\Phi_E$ satisfies a second order differential equation subject to the condition  that $\Phi_E=0$ at the horizon $z=z_h$, we are left with only one constant of integration, which may be either $\mu$ or $\rho$.
Choosing $\mu$ as the integration constant leads to the grand canonical ensemble and the thermodynamic potential density is given by
\begin{align}
\omega(T, \mu)  = -\frac{1}{V\beta}\ln Z = \frac{T}{V}\breve{I}
\end{align}
The variation of $I$ at fixed $\mu$ gives rise to the equations of motion with $I_b=0$ in this case and we can see a straightforward connection between the thermodynamic potential density and Lorentzian Lagrangian:
\begin{align}
  \omega = - \sqrt{|g|} \mathcal{L}_{_\text{matter}}[ -i\Phi_E, \Psi]
\end{align}
On the other hand, the canonical ensemble is obtained by choosing $\rho$ as the integration constant. As we shall see in section 4, a nonzero
$I_b$ is required to cancel the derivative of $\delta\Phi_E$ of $\delta I$, which also
implements the Legendre transformation to the Helmholtz free energy density, i.e.
\begin{align}
\mathfrak{f}(T,\rho) = \omega + \mu \rho = - \sqrt{|\bar g|} \mathcal{L}_{_\text{matter}}[ -i\Phi_E, \Psi]+\frac{T}{V}I_b[\Phi_E,\Psi]
\label{eq:legendre}
\end{align}

The thermodynamic relation
\begin{align}
\bigg(\frac{\partial \omega }{\partial \mu }\bigg)_{T} = \frac{1}{\beta V} \bigg( \frac{\partial \breve{I}}{\partial \mu}\bigg)_{T} = -\rho
\label{thermo}
\end{align}
in the grand canonical ensemble can be easily verified. Consider the variation of $\omega$ under the variation of the integral constant $\mu$, we find
\begin{equation}
\begin{aligned}
\delta\breve{I} &\equiv \beta\int_0^{z_h} \mathrm{d}z\bigg[\delta\Phi_E \frac{\partial \breve{I}}{\partial \Phi_E } + \delta\Psi \frac{\partial \breve{I}}{\partial \Psi }\bigg] \\
& = V\beta\ \bigg[ \frac{\mathrm{d}\Phi_E}{\mathrm{d}z}\ \delta\Phi_E \bigg|_0^{z_h} +  \int_0^{z_h} \mathrm{d}z\ \delta\Phi_E\bigg(- \frac{\mathrm{d}^2\Phi_E  }{\mathrm{d}z^2} + \frac{2\Psi^2 \Phi_E^2}{z^2f }\bigg)  \\
& \qquad \qquad  + 2 \int_0^{z_h} \mathrm{d}z\ \delta\Psi\bigg(- \frac{\mathrm{d}}{\mathrm{d}z}\bigg[\frac{f}{z^2}\frac{\mathrm{d}  }{\mathrm{d}z}\Psi \bigg] - \frac{2\Psi}{z^4} + \frac{\Phi_E^2}{z^2f }\Psi \bigg)\bigg]
\end{aligned}
\end{equation}
where the variations $\delta\Phi_E$ and $\delta\Psi$ are caused by the variation $\delta\mu$ through the equations of motion (\ref{eq:eom-phi}) and (\ref{eq:eom-psi}). The integral
vanishes because of the equations of motion and we are left with
\begin{align}
\delta\breve{I} = V\ \beta\ \Phi_E' \delta\Phi_E\bigg|_0^{z_h} = -V\beta\rho\  \delta\mu.
\end{align}
The relation (\ref{thermo}) follows then.

The Ginzburg-Landau formulation starts with the Euclidean action $I[\Phi_E,\Psi]$ with $\Phi_E = i\Phi$ satisfying (\ref{eq:eom-phi}) but with
$\Psi$ off-shell. The corresponding
thermodynamic potential
\begin{equation}
\omega(T,\mu;\Psi)\equiv \frac{T}{V}I[\Phi_E,\Psi]
\label{eq:omega_GL}
\end{equation}
becomes a functional of $\Psi$. It is straightforward to verify that
\begin{equation}
\bigg(\frac{\partial \omega(T,\mu;\Psi)}{\partial \mu }\bigg)_{T,\Psi} = -\rho
\label{eq:rho-mu}
\end{equation}
with $\rho$ a functional of $\Psi$ now.

\subsection{The order-parameter}
\label{sec-2-3}

The holographic superconductivity involves a correspondence between a hairy black hole in the bulk ($\Psi\neq 0$) and  a long-range order on the boundary. The order parameter is extracted from
the asymptotic form of $\Psi$ towards the boundary. Given the asymptotic
behavior (\ref{asympt:phi}) of $\Phi$, the boundary $z=0$ is a regular point of (\ref{eq:eom-psim}) with indices 1 and 2, we have the asymptotic form:
\begin{align}
 \Psi(z) \approx \Psi_1 [z + O(z^3)] + \Psi_2 z^2 \qquad \text{as } z \rightarrow 0
\end{align}
According to the arguments in Refs.\cite{hartnoll_lectures_2009,horowitz_introduction_2010}, a proper boundary condition amounts to set
either $\Psi_1$ or $\Psi_2$ to zero \footnote{One Lagrangian can give rise to two different quantum field theories in AdS space, depending on the choice of boundary condition, and this quantum field theory in AdS space is equivalent to a conformal field theory on the
 boundary \cite{klebanov_ads/cft_1999}.}, hence the asymptotic behavior becomes \footnote{In the case of $\triangle=1$, a boundary term
$\lim\limits_{z\to 0}\int dt\int d^2\vec x\frac{f}{z^2}\Psi\frac{\partial\Psi}{\partial z}$ should be included in $S_b$ of
(\ref{eq:action-HSC}) to obtain EOM (\ref{eq:eom-psim}) and to render the action finite.}:
\begin{align}
 \Psi(z) \approx \Psi_\triangle z^\triangle, \qquad \text{as } z \rightarrow 0, \quad \text{and} \quad  \triangle = 1 \ \ \text{or } 2.
\label{eq:boundary}
\end{align}
Following the convention in \cite{hartnoll_building_2008}, the order-parameter is defined as:
\begin{align}
 \Braket{\mathcal{O}_\triangle} \equiv \sqrt{2}\ \Psi_\triangle
\end{align}

In the normal phase $\Psi(z)=0$ ($\Psi_1=\Psi_2=0$) and
\begin{equation}
 \Phi=\mu(1-\frac{z}{z_h})=\mu(1-y)
 \label{eq:normal}
\end{equation}
with the charge density $\rho=\frac{\mu}{z_h}$, where $y$ is the scaled coordinate defined at (\ref{eq:matter-action}).
To explore the onset process of the order parameter, we consider a small $\Psi$ such that its feedback to the solution (\ref{eq:normal}) of the $\Phi$-equation can be ignored and
the $\Psi$-equation becomes:
\begin{align}
 \frac{\mathrm{d}  }{\mathrm{d}y}\bigg[(\frac{1}{y^2} - y)\frac{\mathrm{d}  }{\mathrm{d}y} u(y)\bigg] + \frac{2 u(y)}{y^4} + \hat\lambda_\triangle\ \frac{(1 - y)}{y^2(1 + y + y^2)}\ u(y) =0 \label{eq:s-l_psi}
\end{align}
with $\hat\lambda_\triangle=\mu^2\ z_c^2=\rho^2\ z_c^4$.
Eq.(\ref{eq:s-l_psi}) together with boundary condition (\ref{eq:boundary}) and the horizon condition that $\Psi$ is finite at $y=1$ defines a Sturm-Liouville problem \cite{siopsis_analytic_2010}.
A nontrivial solution exists only when $\hat\lambda_\triangle$ coincides with one of the eigenvalues. Let us rank the eigenvalues of the
Sturm-Liouville problem for a given $\triangle$ in an ascending order
$0 < \lambda_0 < \lambda_1 < \lambda_2 < \cdots$ and denote the normalized eigenfunction corresponding the $n$-th eigenvalue by $u_n(y)$. The orthonormal condition reads:
\begin{align}
 \int_0^1 \frac{1 - y}{y^2(1 + y + y^2)} u_m^*(y)u_n(y)\ \mathrm{d}y = \delta_{mn}
\end{align}
The lowest eigenvalue $\lambda_0 > 0$ defines the critical temperature $T_c$ via $\hat\lambda_\triangle=\lambda_0$, i.e.
\begin{equation}
 T_c=\frac{3}{4\pi\sqrt{\lambda_0}}\mu
\end{equation}
In the limit $T\to\infty$, $\hat\lambda_\triangle\to 0$, the only solution is $\Psi=0$ and the system is in its normal phase. As $T$ is lowered to $T_c-0^+$, a nontrivial solution proportional
to $u_0$ develops. The growth of this solution for $T < T_c$ relies on the non-linearity of the equations of motion and the linear Sturm-Liouville analysis
ceases to work. The higher eigenvalues are irrelevant then.

On the other hand, owing to the completeness of eigenfunctions, the solution with $T < T_c$ can be expanded in terms of the orthonormal eigenfunctions $u_i(y)$:
\begin{align}
 \Psi(z) = \sum_{i=0}^\infty \gamma_i(T)\ u_i(y) \label{eq:expansion}
\end{align}
and the order-parameter can be represented as a linear combination of $\gamma_i$'s that are temperature dependent:
\begin{align}
 \Braket{\mathcal{O}_\triangle} =  \sqrt{2}\sum\limits_{i=0}^\infty \gamma_i(T)\ \big(\mathrm{U}_\triangle\big)_i
 \label{eq:order-param}
\end{align}
where $\big(\mathrm{U}_\triangle\big)_i$ is the asymptotic coefficient of the i-th normalized eigenfunction $u_i(y)$:
\begin{align}
 u_i(y) \approx \big(\mathrm{U}_\triangle\big)_i z^\triangle \quad  \text{as} \quad  y \to 0 .
\end{align}

The variational principle underlying the equations of motion is equivalent to the minimization of $\omega(T,\mu;\Psi)$ with respect to $\gamma_i$'s. Solving the minimization problem
conditional to (\ref{eq:expansion}) will lead to a $\braket{O_\triangle}$ dependent thermodynamic potential $\omega(\braket{O_\triangle})\equiv\omega(T,\mu;\Psi)$. As $T\to T_c-0^+$,
only the first term of (\ref{eq:expansion}) and that of (\ref{eq:order-param}) dominate, and we find that:
\begin{align}
 \Psi(y) = \gamma_0\ u_0(y) \qquad \text{as } T \rightarrow T_c.
\end{align}
and $\omega(\braket{O_\triangle})$ takes the Ginzburg-Landau form
\begin{equation}
 \omega(\braket{O_\triangle}) \approx \text{const.}+a\braket{O_\triangle}^2+\frac{1}{2}b\braket{O_\triangle}^4
\end{equation}
with the coefficient $a$ proportional to $T-T_c$. The relation between the order parameter $\braket{O_\triangle}$ and $\gamma_0$ reads:
\begin{align}
 \Braket{\mathcal{O}_\triangle} = \bigg(\frac{4 \pi}{3}\bigg)^\triangle \sqrt{2}\ \mathrm{U}_\triangle  T_c^\triangle \cdot \gamma_0 \qquad \text{as } T \rightarrow T_c
 \label{ODLO}
\end{align}
with $\mathrm{U}_\triangle\equiv (\mathrm{U}_\triangle)_0$.

\section{The Grand Canonical Ensemble}
\label{sec-3}

With the preparation of the proceeding section, it is straightforward to develop the Ginzburg-Landau formulation of a holographic
superconductor and we shall carry out the expedition in the grand canonical ensemble in this section.
The first step is to solve the equation (\ref{eq:eom-phim}) for $\Phi$ in terms of $\Psi$ and substitute the result
into $I[\Phi_E,\Psi]$ of (\ref{eq:euclidean}) to obtain the functional $\omega(T,\mu;\Psi)$ of (\ref{eq:omega_GL}).
Then $\omega[T,\mu;\Psi]$ is expanded to the quartic order in $\Psi$, as is required by the accuracy of the
Ginzburg-Landau theory. The next step is to substitute the expansion (\ref{eq:expansion}) into $I[\Phi_E, \Psi]$, which turn
it into a function of infinite variables, $\gamma_j$'s. Minimizing $\omega(T,\mu; \Psi)$ with
respect to $\gamma_j$'s with $j>0$ yields $\gamma_j=O(\gamma_0^3)$ for a small $\gamma_0$ and thereby
$\braket{\mathcal{O}_\triangle}=O(\gamma_0)$. Up to the quartic order in $\gamma_0$, $\gamma_j$ with $j>0$ may be ignored and the resultant
$\omega[T,\mu;\Psi]$ (a quartic polynomial on $\gamma_0$) is the Ginzburg-Landau energy we are pursuing.

We shall consider the order parameter with $\triangle= 1, \text{or}\  2$, respectively. For a specified conformal dimension $\triangle$:
\begin{equation}
 \Psi(z) = \Psi_\triangle z^\triangle +O(z^3) \qquad \text{as} \qquad z\to 0.
 \label{bdry:psi}
\end{equation}

An arbitrary variation $\delta\Psi(z)$ subject to the conditions $\delta\Psi(z)\to\delta\Psi_\triangle z^\triangle$ as $z\to 0$ and
$\delta\Psi(z_h)=0$ leads to the equation (\ref{eq:eom-psim}) with $I_b=0$.

The $\Phi$ field equation (\ref{eq:eom-phim}) can be converted to an integral equation:
\begin{align}
 \Phi(z) = \mu -\rho\ z - 2 \int_0^z \frac{\Psi^2(\zeta)}{\zeta f(\zeta)}\Phi(\zeta)\ \mathrm{d}\zeta
         + 2z \int_0^z \frac{\Psi^2(\zeta)}{\zeta^2 f(\zeta)}\Phi(\zeta)\ \mathrm{d}\zeta
 \label{integral1}
\end{align}
which can be solved iteratively in powers of $\Psi$. In the normal phase $\Psi=0$ and we have an exact solution $\Phi=\mu - \rho z$ with
the density $\rho=\frac{\mu}{z_h}$.
In the super phase, the asymptotic condition (\ref{bdry:psi})
ensures that the integrals in (\ref{integral1}) are of the order $O(z^4)$ and therefore the asymptotic form (\ref{asympt:phi})
can be isolated explicitly. The density $\rho$ as a function of $\mu$ and $T$ is fixed by the condition $\Phi(z_h) \equiv 0$ and we find that
\begin{align}
 \rho = \frac{\mu}{z_h} + 2\int_0^{z_h} J \Psi^2\ \Phi\ \mathrm{d}z  \label{eq:rho}
\end{align}
where $J(z) \equiv \dfrac{1 - \frac{z}{z_h}}{z^2(1 - \frac{z^3}{z_h^3})}=\dfrac{1}{z_h^2y^2(1+y+y^2)}$ for brevity. At this stage, $\rho$ is also a functional of $\Psi$.
Substituting (\ref{eq:rho}) into the (\ref{integral1}), the integral equation becomes:
\begin{align}
 \Phi(z) = \mu(1 - \frac{z}{z_h}) - 2(1 - \frac{z}{z_h})\int_0^z \frac{\Psi^2}{\zeta f}\Phi\ \mathrm{d}\zeta
  - 2 z \int_z^{z_h}J\ \Psi^2\ \Phi\ \mathrm{d}\zeta
 \label{integral2}
\end{align}
Meanwhile, the equation of motion (\ref{eq:eom-phim}) enables us to write
\begin{align}
 \omega(T, \mu ;\Psi) = -\frac{1}{2}\mu \rho + \int_0^{z_h} \mathrm{d}z \bigg[ \frac{f}{z^2}\big(\frac{\mathrm{d}\Psi}{\mathrm{d}z})^2 - \frac{2\Psi^2}{z^4}\bigg] \label{eq:omega}
\end{align}
where the first term on right hand side comes from the integration by part of the derivative term
$\big(\dfrac{\mathrm{d}\Phi}{\mathrm{d}z}\big)^2$ in the action (\ref{eq:euclidean}).

We need $\Phi$ to the quadratic order of $\Psi$ in order to expand
the thermodynamic potential to the quartic order in $\Psi^4$.
Starting with the leading order of $\Phi$ field:
\begin{align}
 \Phi(z) = \mu\bigg(1 - \frac{z}{z_h}\bigg)+O(\Psi^2) \label{eq:phi_0}
\end{align}
we find:
\begin{equation}
 \Phi(z) = \mu\bigg(1 - \frac{z}{z_h}\bigg) - 2\mu\bigg(1 - \frac{z}{z_h}\bigg)
 \int_0^z \zeta\ J\Psi^2\ \mathrm{d}\zeta
 - 2\mu z\int_z^{z_h} \left(1-\frac{\zeta}{z_h}\right)J\Psi^2\ \mathrm{d}\zeta + O(\Psi^4)
 \label{eq:phi_2}
\end{equation}
and
\begin{equation}
\begin{aligned}
 \rho = \frac{\mu}{z_h} + 2\mu \int_0^{z_h} & \left(1- \frac{z}{z_h}\right)J \Psi^2\ \mathrm{d}z \\
&-8 \mu \int_0^{z_h} \mathrm{d}z \bigg(z\ J(z) \Psi^2(z)
 \int_z^{z_h} \left(1-\frac{\zeta}{z_h}\right)J(\zeta) \Psi^2(\zeta)\ \mathrm{d}\zeta \bigg) + O\big(\Psi^6\big )\label{eq:rho_2}
\end{aligned}
\end{equation}

Substituting (\ref{eq:rho_2}) into (\ref{eq:omega}), we obtain that:
\begin{equation}
 \begin{aligned}
   \omega(T, \mu; \Psi) &=  -\frac{\mu^2}{2z_h} + \int_0^{z_h} \mathrm{d}z \Bigg[ \frac{f}{z^2}\big( \frac{\mathrm{d} \Psi}{\mathrm{d}z}  \big)^2 - \frac{2\Psi^2}{z^4} - \mu^2\left(1-\frac{z}{z_h}\right)J\Psi^2  \Bigg]\\
   &\hspace{2cm} + \int_0^{z_h} \mathrm{d}z \bigg\{4 \mu^2\left(1-\frac{z}{z_h}\right)J(z)\Psi^2(z) \int_0^z J(\zeta)\ \zeta\ \Psi^2(\zeta) \ \mathrm{d}\zeta\bigg\} + O\big(\Psi^6\big)  \\
   &= -\frac{\mu^2}{2z_h} + \frac{1}{z_h^3}\int_0^1 \mathrm{d}y\ \bigg[
   \Psi\bigg( -\frac{\mathrm{d}}{\mathrm{d}y} \left(\frac{1-y^3}{y^2}\right)\frac{\mathrm{d}}{\mathrm{d}y} - \frac{2}{y^4} - \hat\lambda_\triangle\ \frac{1-y}{y^2(1+y+y^2)}\bigg) \Psi \bigg]\\
   &\hspace{0.3cm} +  4\ \frac{\hat\lambda_\triangle}{z_h^3}\int_0^1 \mathrm{d}y\ \bigg\{ \frac{1-y}{y^2(1+y+y^2)}\Psi^2(y) \int_0^y\mathrm{d}\eta\ \frac{1}{\eta(1+\eta+\eta^2)}\Psi^2(\eta) \bigg\} + O(\Psi^6)\label{eq:omega_4}
 \end{aligned}
\end{equation}
where in the last step an integration by part is made and the final expression is written in terms of the scaled coordinate $y$
defined at (\ref{eq:matter-action}) with $\hat\lambda_\triangle = \mu^2 z_c^2$. We recognize the Sturm-Liouville operator
sandwiched between two $\Psi$'s in the first term inside the bracket of the final expression of (\ref{eq:omega_4}). We also notice that the
quartic term of (\ref{eq:omega_4}) is always positive for a nonzero $\Psi$.
Similarly, the equation (\ref{eq:eom-psim}) becomes an integro-differential equation:
\begin{equation}
 \begin{aligned}
   \frac{\mathrm{d}  }{\mathrm{d}y}\bigg[\left(\frac{1-y^3}{y^2}\right) \frac{\mathrm{d}  }{\mathrm{d}y}\Psi \bigg] & + \frac{2\Psi}{y^4} + \hat\lambda_\triangle\ \frac{1 - y}{y^2(1 + y +y^2)} \Psi =\\
   & \quad \ 4\hat\lambda_\triangle \frac{1 - y}{y^2(1 + y +y^2)}\Psi(y) \int_0^y \frac{1}{\eta(1+\eta+\eta^2)} \Psi^2(\eta)\ \mathrm{d}\eta \\
   & \qquad+ 4\hat\lambda_\triangle \frac{1}{y(1 + y + y^2)}\Psi(y) \int_y^{1} \frac{1 -\eta }{\eta^2(1 + \eta + \eta^2)} \Psi^2(\eta)\ \mathrm{d}\eta
+ O(\Psi^5) \label{eq:psi_gce}
 \end{aligned}
\end{equation}
Upon substitution of (\ref{eq:expansion}), we find that
\begin{equation}
 \omega(T, \mu; \Psi)=-\frac{\mu^2}{2z_h}+\frac{1}{z_h^3}\sum_{n=0}^\infty(\lambda_n-\hat\lambda_\triangle)\gamma_n^2  + \frac{2}{z_h^3}\sum_{i,j,k,l=0}^\infty A_{ijkl}\gamma_i\gamma_j\gamma_k\gamma_l,
 \label{eq:omega_g}
\end{equation}
where
\begin{equation}
 \begin{aligned}
   A_{ijkl} = &\int_0^1 \mathrm{d}y \bigg( \frac{1 -y }{y^2(1 + y + y^2)}\ u_i(y)u_j(y) \int_0^y \frac{1}{\eta(1 + \eta + \eta^2)}\ u_k(\eta)u_l(\eta)\ \mathrm{d}\eta \bigg) \\
   &+ \int_0^1 \mathrm{d}y \bigg(\frac{1}{y(1 + y + y^2)}\ u_i(y)u_j(y) \int_y^1 \frac{1 -\eta }{\eta^2(1 + \eta + \eta^2)}\ u_k(\eta)u_l(\eta)\ \mathrm{d}\eta \bigg)
 \end{aligned}
\end{equation}
and is symmetric with all subscripts.
In particular,
\begin{align}
 A_{0000} = 2 \int_0^1 \mathrm{d}y\bigg(\frac{1}{y(1 + y + y^2)}\  u_0^2(y) \int_y^1 \frac{1 -\eta }{\eta^2(1 + \eta + \eta^2)}\ u_0^2(\eta) \mathrm{d}\eta \bigg) \equiv 2\ \mathcal{G}_\triangle.
\end{align}
As $T\to T_c$ from below, $\hat\lambda_\triangle\to \lambda_0$, the coefficient of $\gamma_0^2$ goes to zero from below while the coefficients of other $\gamma$'s remains positive and O(1).
Minimizing (\ref{eq:omega_g}) with respect to {\it all} $\gamma$'s is equivalent to solving the equation (\ref{eq:psi_gce}).
For the purpose of the Ginzburg-Landau formulation, however, we minimize (\ref{eq:omega_g}) with respect $\gamma_n$ with
$n\ge 1$ conditional on a small $\gamma_0$. We find:
\begin{align}
 \bigg(1 - \frac{\lambda_n}{\hat\lambda_\triangle}\bigg) \gamma_n = 4\sum_{i,j,k =0}^\infty \gamma_i \gamma_j \gamma_k \cdot A_{nijk}   \label{eq:gamma-gce}
%\label{conditional}
\end{align}
with $n\ge 1$. Since the coefficient on left hand side is of the order O(1) and the leading term on right hand side is
proportional to $\gamma_0^3$, we have
\begin{equation}
 \gamma_n = O(\gamma_0^3)
\label{eq:gamma_n}
\end{equation}
for $n\ge 1$. It follows that all $\gamma_n$'s with $n\ge 1$ in (\ref{eq:omega_g}) and in the expression of $\braket{\mathcal{O}_\triangle}$ of (\ref{eq:order-param}) can be ignored and
the Ginzburg-Landau form of $\omega[T,\mu;\Psi]$ emerges:
\begin{equation}
 \begin{aligned}
 \omega(T, \mu; \Psi) &= -\frac{\mu^2}{2z_h}+\frac{1}{z_h^3}(\hat\mu_c^2-\hat\lambda_\triangle)\gamma_0^2+\frac{4\mathcal{G}_\triangle\mu^2}{z_h}\gamma_0^4  \\
 & \approx -\frac{\mu^2}{2 z_h} + a_{_\text{GCE}} \braket{\mathcal{O}_\triangle
}^2+\frac{1}{2}b_{_\text{GCE}} \braket{\mathcal{O}_\triangle}^4
 \label{eq:GL_gce}
 \end{aligned}
\end{equation}
where
\begin{equation}
\begin{aligned} \label{eq:3}
  a_{_\text{GCE}} &=  \frac{\lambda_0}{\mathrm{U}_\triangle^2}\ \left(\frac{4 \pi}{3} T_c\right)^{3 - 2\triangle}\  \bigg(\frac{T}{T_c}-1\bigg)  \\
  b_{_\text{GCE}} &= 2\ \mathcal{G}_\triangle\ \frac{\lambda_0\ }{\mathrm{U}_\triangle^4} \left(\frac{4 \pi}{3} T_c\right)^{3 - 4\triangle}
\end{aligned}
\end{equation}
with the critical temperature
\begin{equation}
 T_c  = \frac{3\mu}{4\pi\sqrt{\lambda_0}}.
\end{equation}
Minimize (\ref{eq:GL_gce}) with respect to $\braket{\mathcal{O}_\triangle}$, we obtain that:
\begin{equation}
  \Braket{\mathcal{O}_\triangle} = \left(\frac{4 \pi}{3}\right)^\triangle \ \frac{\mathrm{U}_\triangle}{\sqrt{2 \mathcal{G}_\triangle}}\ T_c^\triangle\sqrt{1 - \frac{T}{T_c}}
\label{eq:critical-GCE}
\end{equation}
and the on-shell thermodynamic potential density:
\begin{equation}
 \omega(T,\mu)  = -\frac{\mu^2}{2 z_h} - \frac{\mu^2}{4 z_h\mathcal{G}_\triangle} \bigg(1 - \frac{T}{T_c} \bigg)^2.
\end{equation}

The Sturm-Liouville equation (\ref{eq:s-l_psi}) is a Fuchs equation of five regular points and cannot be solved analytically. For the conformal dimension two case, $\triangle =2$,  Our numerical solution with the 4th-order Runge-Kutta method yields :

\begin{align}
  \lambda_0 = \hat\lambda_2 = 16.515 ; \quad \mathcal{G}_2 = 0.3125; \quad  \mathrm{U}_2 =  7.375
\label{data-2}
\end{align}
Consequently, the critical temperature
\begin{equation}
 T_c \approx 0.0587\; \mu,
\end{equation}
the Ginzburg-Landau coefficients
\footnote{
Our GL coefficients,  in $\triangle =1$ and $2$ below, are different from the ones reported in \cite{herzog}, for $\triangle =2$, their numerical results are
$a_{_\text{GCE}}=\frac{5.13}{\mu}\left(\frac{T}{T_c}-1\right)$ and $b_{_\text{GCE}}=\frac{2.68}{\mu^5}$, in the absence of vorticity.
However,  the straightforward Legendre transformation, (\ref{eq:legendre}) and (\ref{thermo}), leads to the scaling
law $\braket{\mathcal{O}_2}=88.8T_c^2\sqrt{1-\frac{T}{T_c}}$ in the canonical ensemble with the coefficient different from that
reported in \cite{hartnoll_building_2008}. This inconsistency also happens in $\triangle=1$, which their GL coefficients are $a_{_\text{GCE}}=3.07\mu\left(\frac{T}{T_c}-1\right)$ , $b_{_\text{GCE}}=\frac{1.486}{\mu}$, and the corresponding scaling law in the canonical ensemble is $\braket{\mathcal{O}_1}=3.45 T_c\sqrt{1 - \frac{T}{T_c}}$ .}:
\begin{equation}
\begin{aligned}
  a_{_\text{GCE}} &\approx 0.0725\ \frac{1}{T_c}\bigg(\frac{T}{T_c}-1\bigg) \approx \frac{1.24}{\mu}\bigg(\frac{T}{T_c}-1\bigg),   \\
  b_{_\text{GCE}} &\approx 2.706 \times 10^{-6}\ \frac{1}{T_c^5} \approx \frac{3.88}{\mu^5},
\end{aligned}
\end{equation}
the critical scaling of the order parameter
\begin{align}
  \Braket{\mathcal{O}_2} = 163.68\ T_c^2\sqrt{1 - \frac{T}{T_c}}
\label{eq:163-GCE}
\end{align}
and the on-shell thermodynamic potential
\begin{equation}
\begin{aligned}
 \omega(T,\mu) \approx  - 606.896\ T_c^3 - 971.034\ T_c(T_c - T)^2.
\end{aligned}
\end{equation}

Following the method above with a modified boundary condition of the Sturm-Liouville problem
at $z=0$, it's straightforward to switch into the conformal dimension one case, $\triangle = 1$, we obtain:
\begin{align}
  \lambda_0 = \hat\lambda_1 = 1.241; \quad   \mathrm{U}_1 = 1.75;\ \quad  \mathcal{G}_1 = 0.1003; \quad  T_c = 0.214\ \mu
\label{data-1}
\end{align}
Then the formula (\ref{eq:critical-GCE}) gives rise to the order-parameter:
\begin{align}
  \braket{\mathcal{O}_1} =  16.37\ T_c\sqrt{1 - \frac{T}{T_c}}
\end{align}
from (\ref{eq:3}), we can obtain the Ginzburg-Landau coefficient $a_{_\text{GCE}}$ and $b_{_\text{GCE}}$,
\begin{equation}
\begin{aligned}
  a_{_\text{GCE}} &= 1.696\ (T - T_c)= 0.363\ \mu\left(\frac{T}{T_c} - 1\right) \\
  b_{_\text{GCE}} &= 6.33\times 10^{-3}\ \frac{1}{T_c} = 0.0296\  \frac{1}{\mu}
\end{aligned}
\end{equation}
and the on-shell Ginzburg-Landau free energy:
\begin{align}
  \omega = - 45.568\ T_c^3 - 227.204\ T_c(T_c - T)^2
\end{align}

\section{The Canonical Ensemble}
\label{sec-4}

The thermodynamic variables in the canonical ensemble are the temperature and the charge density. The characteristic thermodynamic function is the Helmholtz free energy,
which is a Legendre transformation of the thermodynamic potential in the previous section, i.e.
\begin{align}
 \mathfrak{f}(T, \rho) = \omega(T, \mu) + \mu\rho.
 \label{eq:f_onshell}
\end{align}
For the sake of the Ginzburg-Landau formulation, we shall work with a Helmholtz free energy with an off-shell $\Psi$, i.e.
\begin{equation}
 \mathfrak{f}(T,\rho;\Psi) = \omega(T,\mu;\Psi)+\mu\rho
 \label{eq:f_psi}
\end{equation}
Therefore we need a functional relation, $\mu = \mu(T,\rho,\Psi)$, to convert $\mu$ on the right hand side of (\ref{eq:f_psi}) to $\rho$ and this can be
obtained from (\ref{eq:rho-mu}), or equivalently, from (\ref{eq:rho_2}). Two approaches leading to the same
Ginzburg-Landau formulation in the canonical ensemble are followed in this section.

Our first approach is purely thermodynamic. Starting with the Ginzburg-Landau form of the
thermodynamic potential (\ref{thermo}) in the grand canonical potential, Eq.(\ref{eq:rho-mu}) yields
\begin{equation}
 \rho = \frac{\mu}{z_h}\bigg[1 + 2\ \gamma_0^2 - 8\ \mathcal{G}_\triangle\gamma_0^4 + O(\gamma_0^6)\bigg]
\end{equation}
and it follows that
\begin{equation}
 \mu = \rho z_h \bigg[1-2\gamma_0^2+4(2\mathcal{G}_\triangle + 1)\gamma_0^4+O(\gamma_0^6)\bigg]
 \label{eq:mu_rho}
\end{equation}
Consequently
\begin{equation}
 \begin{aligned}
 \mathfrak{f}(T,\rho;\Psi) &= \frac{1}{2}\rho^2z_h + \frac{1}{z_h^3}(\lambda_0-\rho^2z_h^4)\gamma_0^2 + 2 \rho^2 z_h (2\mathcal{G}_\triangle + 1)\gamma_0^4 + O(\gamma_0^6)\\
 &= \frac{1}{2}\rho^2z_h + a_{_\text{CE}}\braket{\mathcal{O}_\triangle}^2 + \frac{1}{2}b_{_\text{CE}}\braket{\mathcal{O}_\triangle}^4 + O(\braket{\mathcal{O}_\triangle}^6),
 \label{eq:f_gamma}
\end{aligned}
\end{equation}
with
\begin{equation}
\begin{aligned}
  a_{_\text{CE}} &= 2\ \frac{\lambda_0}{U_\triangle^2}\ \left(\frac{4\pi}{3}\ T_c\right)^{3 - 2\triangle}\ \left( \frac{T}{T_c} -1 \right)  \\
  b_{_\text{CE}} &= \frac{\lambda_0}{U_\triangle^4}\left(2 \mathcal{G}_\triangle + 1\right)  \left(\frac{4\pi}{3}\ T_c\right)^{3 - 4\triangle}
  \label{eq:coeff-CE}
\end{aligned}
\end{equation}
Minimizing (\ref{eq:f_gamma}) with respect $\Braket{\mathcal{O}_\triangle}$, we obtain the critical behavior of the order parameter
\begin{align}
  \Braket{\mathcal{O}_\triangle} = \left(\frac{4 \pi}{3}\right)^\triangle \mathrm{U}_\triangle \sqrt{\frac{2}{2 \mathcal{G}_\triangle + 1}} \cdot T_c^\triangle \sqrt{1 - \frac{T}{T_c}}
  \label{CE:order-para}
\end{align}
As a result, the on-shell Ginzburg-Landau free energy is:
\begin{align}
  \mathfrak{f} = \frac{1}{2}\left(\frac{4\pi}{3}\right)^3 \lambda_0\ T_c^3 - 2\ \left(\frac{4 \pi}{3}\right)^3 \lambda_0 \frac{1}{2 \mathcal{G}_\triangle +1}\ T_c(T_c - T)^2
\label{eq:4}
\end{align}

For the order parameter of conformal dimension two, $\triangle = 2$, we have the critical temperature
\begin{equation}
 T_c = \frac{3}{4\pi}\hat\lambda_2^{-\frac{1}{4}}\sqrt{\rho} = 0.118\ \sqrt{\rho}.
\end{equation}
and the Ginzburg-Landau coefficients :
\begin{equation}
\begin{aligned}
  a_{_\text{CE}} &= 0.145\ \frac{1}{T_c}\left(\frac{T}{T_c} - 1\right)   \\
  b_{_\text{CE}} &=  7.0346 \times 10^{-6}\ \frac{1}{T_c^5},
\end{aligned}
\end{equation}
It follows from (\ref{data-2}) and (\ref{CE:order-para}), the critical behavior of the order parameter reads:
\begin{align}
  \Braket{\mathcal{O}_2} = 143.574\ T_c^2 \sqrt{1 - \frac{T}{T_c}}
\label{eq:144-CE}
\end{align}
and the on-shell Ginzburg-Landau free energy in the canonical ensemble becomes:
\begin{equation}
 \begin{aligned}
   \mathfrak{f}(T, \rho) =  606.896\ T_c^3 - 1493.898\ T_c(T_c - T)^2. \label{eq:c.e.}
 \end{aligned}
\end{equation}

In the case of the conformal dimension one, $\triangle=1$, we have $T_c \approx 0.226\ \sqrt{\rho}$ and the
off-shell Ginzburg-Landau free energy
\begin{equation}
\begin{aligned}
  \mathfrak{f} &= \frac{1}{2}\rho^2 z_h + a_{_\text{CE}} \braket{O_1}^2 + \frac{1}{2}b_{_\text{CE}}\braket{O_1}^4 \\
  & = \frac{1}{2}\rho^2 z_h + 3.391\ (T-T_c) \ \braket{O_1}^2 + 1.894 \times 10^{-2}\ \frac{1}{T_c} \braket{O_1}^4
\label{GLO1}
\end{aligned}
\end{equation}

It follows that
\begin{align}\label{eq:1}
  \braket{O_1} = 9.462\ T_c \sqrt{1 - \frac{T}{T_c}}
\end{align}
and the on-shell form of (\ref{GLO1}) is given by
\begin{align}
  \mathfrak{f} = 45.568\ T_c^3 - 512.398\ T_c(T_c - T)^2
\end{align}

The values of the scaling coefficients in order-parameter (\ref{eq:144-CE});  (\ref{eq:1}) and critical temperature relations make a good agreement with that reported in \cite{hartnoll_building_2008} , which is obtained by numerical fitting.

It is interesting to notice that the quartic terms of the Ginzburg-Landau energy in the grand canonical ensemble and in the canonical ensemble are not
simply related by the leading order relation $\frac{\mu^2}{z_h} = \rho^2z_h$ and this leads to different coefficients of the
critical behaviors (\ref{eq:163-GCE}) and (\ref{eq:144-CE}) in the two ensembles.
The difference stems from the contribution of the $O(\gamma_0^2)$ term of the
expansion (\ref{eq:mu_rho}) in the Legendre transformation to the quartic term of $\mathfrak{f}$. The same
Legendre transformation for a BCS superconductor is reproduced in the appendix. There we see that the  contribution of the quadratic term in the expansion of $\mu$ in the powers of the order parameter to the quartic term of the Helmholtz free energy
is suppressed because of $T_c \ll\mu$.
This observation reflects the strong coupling nature of the holographic superconductivity.

Our second approach highlights the $\Phi_E$-dependent term of the boundary action $I_b$ necessary to produce the equation of motion (\ref{eq:eom-phi})
and to implement the Legendre transformation (\ref{eq:f_psi}). A similar mechanism \footnote{A specified boundary condition for the gravitational action corresponds to a Legendre transformation.} was discussed in the context of the black hole
thermodynamics in Refs: \cite{brown_thermodynamic_1990,braden_charged_1990}. The integration constant for the equation of motion of $\Phi_E$ is the density $\rho$ now
and the variation $\delta\Phi_E$ underlying (\ref{eq:eom-phi}) satisfies the condition
\begin{equation}
 \delta\frac{\mathrm{d}\Phi_E}{\mathrm{d}z}\to 0 \qquad \delta\Phi_E \to \delta\mu\neq 0
\end{equation}
at the boundary $z=0$ in this case. Therefore a term
\begin{equation}
 I_b = V\beta\Phi_E \frac{\mathrm{d}\Phi_E}{\mathrm{d}z}\bigg|_0^{z_h} = V\beta\mu\rho
\end{equation}
is required such that the boundary term of $\delta I[\Phi_E,\Psi]$ from the integration by part is canceled by $\delta I_b$. This additional term provides
precisely the $\mu\rho$ term of the Legendre transformation from the grand canonical ensemble to the canonical ensemble.
It follows from (\ref{eq:rho-mu}) that
\begin{equation}
 \begin{aligned}
   \mu &= \rho z_h \Bigg[1 - 2 \int_0^{z_h} \left(1-\frac{z}{z_h}\right)J \Psi^2\ \mathrm{d}z + 4 \left(\int_0^{z_h} \left(1-\frac{z}{z_h}\right)J \Psi^2\ \mathrm{d}z\right)^2 \\
   & \qquad + 8  \int_0^{z_h} \mathrm{d}z \left\{ z J(z) \Psi^2(z) \int_z^{z_h} \left(1-\frac{\zeta}{z_h}\right)J(\zeta) \Psi^2(\zeta)\ \mathrm{d}\zeta \right\} \Bigg] + O(\Psi^6)
 \end{aligned}
\end{equation}
for an arbitrary $\Psi$ with the asymptotic form (\ref{bdry:psi}). Substituting it into (\ref{eq:f_psi}), we obtain that:
\begin{equation}
 \begin{aligned}
   \mathfrak{f}(T,\rho;\Psi) &= \frac{1}{2}\rho^2 z_h + \int_0^{z_h} \mathrm{d}z\ \Psi\bigg[ -\frac{\mathrm{d}}{\mathrm{d}z}\frac{f}{z^2}
   \frac{\mathrm{d}}{\mathrm{d}z} - \frac{2}{z^4} - \rho^2 z_h^2 \left(1-\frac{z}{z_h}\right)J\bigg] \Psi \\
   & \qquad  + 4 \rho^2 z_h^2 \int_0^{z_h} \mathrm{d}z \left\{ z J(z) \Psi^2(z) \int_z^{z_h} \left(1-\frac{\zeta}{z_h}\right)J(\zeta) \Psi^2(\zeta)\  \mathrm{d}\zeta \right\}   \\
   & \qquad \quad  \hrulefill + 2 \rho^2 z_h^3 \bigg[\int_0^{z_h} \left(1-\frac{z}{z_h}\right)J \Psi^2\ \mathrm{d}z\bigg]^2
   \label{eq:frakf}
\end{aligned}
 \end{equation}
We notice that the quartic term is different from that of (\ref{eq:omega}) but remains positive for a nonzero $\Psi$. Substituting the
expansion (\ref{eq:expansion}) into (\ref{eq:frakf}),
we find that $\Psi(z)$ is dominated by the leading term $i=0$ as $T\to T_c$ and the same Ginzburg-Landau free energy (\ref{eq:c.e.}) emerges.

 \section{The Gradient Term}
 \label{sec-5}

The main application of the Ginzburg-Landau theory is to describe an inhomogenous condensate controlled by the gradient term
in the thermodynamic potential or free energy. Let us discuss this term in the context of a holographic superconductor in the probe limit. The gradient here refers to the gradient with respect to the transverse coordinates $\vec x=(x_1, x_2)$.
With the $\vec x$-dependence, $\Phi = \Phi(z,\vec x); \ \Psi = \Psi(z,\vec x)$, the action (\ref{eq:action}) takes the form
\begin{equation}
S_{_\text{matter}} = \int \mathrm{d}t\int \mathrm{d}^2 \vec x\int_0^{z_h} \mathrm{d}z\ \sqrt{-g}\
\mathcal{L}_{_\text{matter}}[\Phi,\Psi]
\label{eq:action-x}
\end{equation}
with
\begin{equation}
\mathcal{L}_{_\text{matter}}[\Psi,\Phi] =  \frac{1}{2f}(\vec\nabla\Phi)^2-\frac{1}{z^2}|
\vec\nabla\Psi|^2+\frac{1}{2} \big|\frac{\partial \Phi}{\partial z}\big|^2 - \frac{f}{z^2} \big|\frac{\partial \Psi}{\partial z}\big|^2
+ \frac{|\Psi|^2\Phi^2}{z^2f} + \frac{2|\Psi|^2}{z^4},
\end{equation}
where $\vec\nabla=(\frac{\partial}{\partial x_1},\frac{\partial}{\partial x_2})$ and $\Psi$ is allowed to be complex. In Lorentzian formulation ,the equation of motion for the Maxwell field $\Phi(z, \vec x)$, (\ref{eq:eom-phi}), is replaced by
\begin{align}
 \frac{\partial^2 \Phi}{\partial z^2 } = \frac{2|\Psi|^2}{z^2f}\Phi - \frac{1}{f}\vec\nabla^2\Phi\label{eq:phi-x}
\end{align}
We shall show now that with $\Phi$ satisfying (\ref{eq:phi-x}), the gradient term of $\Phi$ in (\ref{eq:action-x}) has no contribution to the order of the Ginzburg-Landau theory.

Let us consider first the grand canonical ensemble. Subject to the requirement $\Phi(z_h,\vec{x})= 0$ and the AdS boundary condition $\Phi(z,\vec{x})\bigg|_{z=0} = \mu$, with $\mu,\vec x$-independent, the integral form of (\ref{eq:phi-x}) reads:
  \begin{equation}
    \begin{aligned}
      \Phi(z,\vec{x}) &= \mu(1 - \frac{z}{z_h}) -2(1 - \frac{z}{z_h})\int_0^z \frac{|\Psi(\zeta,\vec{x})|^2}{\zeta f(\zeta)}\Phi(\zeta,\vec{x})\ \mathrm{d}\zeta - 2\ z\int_z^{z_h} J(\zeta)|\Psi(\zeta,\vec{x})|^2 \Phi(\zeta,\vec{x}) \mathrm{d}\zeta \\
      & \hspace{1cm}  + \int_0^z \frac{\zeta \vec\nabla^2 \Phi(\zeta,\vec{x})}{f(\zeta)} \mathrm{d}\zeta - z\int_0^z \frac{\vec\nabla^2 \Phi(\zeta,\vec{x})}{f(\zeta)} \mathrm{d}\zeta
      - \int_0^{z_h} \zeta^2J(\zeta)\vec\nabla^2\Phi(\zeta,\vec{x}) \mathrm{d}\zeta \label{eq:phi-x-integral}
      \end{aligned}
    \end{equation}
Correspondingly we have
    \begin{align}
      \rho = \frac{\mu}{z_h} + 2\int_0^{z_h} J |\Psi|^2\ \Phi\ \mathrm{d}z -2\int_0^{z_h} z^2J\ \vec\nabla^2\Phi\ \mathrm{d}z    \label{eq:density}
    \end{align}
which is $\vec x$-dependent through $\Psi$, and the thermodynamic potential becomes:
\begin{equation}
    \begin{aligned}
      \Omega(T,\mu;\Psi) = TI[\Phi_E,\Psi]
      &= -\int \mathrm{d}^2 \vec x\int_0^{z_h} \mathrm{d}z\ \sqrt{g_{_E}}\
\mathcal{L}[-i\Phi_E,\Psi]  \\
       &= \int \mathrm{d}^2 \vec x\bigg[ -\frac{1}{2}\mu \rho + \int_0^{z_h} \mathrm{d}z
      \bigg(\frac{1}{z^2}|\vec\nabla\Psi|^2 + \frac{f}{z^2}|\frac{\mathrm{d}\Psi}{\mathrm{d}z}|^2 - \frac{2|\Psi|^2}{z^4}\bigg)\bigg]   \label{eq:Omega}
    \end{aligned}
\end{equation}
    It is noteworthy that the Laplacian of $\Phi$ in (\ref{eq:density}) makes no contribution when (\ref{eq:density}) is
substituted into contribution (\ref{eq:Omega}).

Employing the same iterative procedure of the section 3, we find the leading order of the Maxwell field $\Phi$ has the same form as (\ref{eq:phi_0}):
\begin{align}
 \Phi(z,\vec{x}) = \mu(1 - \frac{z}{z_h}) + O(\Psi^2)
\end{align}
which is independent of $\vec x$.
Consequently, the $\vec x$-dependence of $\Phi$ stems from the $\vec x$-dependence of $\Psi$ in the next order of the iteration, i.e.
\begin{equation}
\begin{aligned}
  \Phi(z,\vec{x}) &= \mu\bigg(1 - \frac{z}{z_h}\bigg) - 2\mu\bigg(1 - \frac{z}{z_h}\bigg) \int_0^z \zeta J(\zeta) |\Psi(s,\vec{x})|^2\mathrm{d}\zeta \\
& \hspace{4cm}- 2\mu\ z\int_z^{z_h} \left(1-\frac{\zeta}{z_h}\right)J(\zeta)|\Psi(\zeta,\vec{x})|^2\mathrm{d}\zeta + O(\Psi^4).
  % \label{eq:phi-x_2}
\label{eq:Phi_2-x}
\end{aligned}
\end{equation}
It follows that
\begin{equation}
\vec\nabla\Phi = O(\vec\nabla|\Psi|^2).
\end{equation}
Consequently, the contribution from the gradient square,
$(\vec\nabla\Phi)^2$ of (\ref{eq:action-x}) is beyond the order of the Ginzburg-Landau gradient term, $|\vec\nabla\Psi|^2$, and can be dropped.
Therefore the gradient term of the order parameter within GL framework comes solely from the second term
of (\ref{eq:action-x}).

Substituting (\ref{eq:Phi_2-x}) and (\ref{eq:density}) together with the expansion the expansion
\begin{align}
 \Psi(z,\vec{x}) = \sum\limits_{i=_0}^\infty \gamma_i(\vec{x})\ u_i(z)
\end{align}
into (\ref{eq:Omega}), we obtain that:
\begin{equation}
\begin{aligned}
 \Omega[T,\mu;\Psi] &= \Omega[T,\mu;0] + \frac{1}{z_h^3}\int \bigg[\sum_{i,j=0}^\infty Q_{ij} \vec\nabla\gamma_i^*\cdot \vec\nabla\gamma_j + \sum\limits_{n=_0}^\infty(\lambda_n-\hat\lambda_\triangle)|\gamma_n|^2\ \bigg] \mathrm{d}^2\vec x \\
&\hspace{6cm}  + \frac{2}{z_h^3}\sum_{i,j,k,l=0}^\infty\int \mathrm{d}^2\vec x\ A_{ijkl}\gamma_i^*\gamma_j\gamma_k^*\gamma_l
\end{aligned}
\end{equation}
where
\begin{align}
 Q_{i j} \equiv \int_0^1 \frac{u_i u_j}{y^2}\ \mathrm{d}y
\end{align}
and the eq.(\ref{eq:gamma-gce}) for $n\ge 1$ becomes
\begin{align}
 \bigg( -\vec\nabla^2 + \frac{\lambda_n}{\hat\lambda_\triangle}-1\bigg) \gamma_n = \sum\limits_{i\ge 1}^\infty  Q_{n i} \vec\nabla^2 \gamma_i - 4\sum_{i,j,k =0}^\infty \gamma_i \gamma_j^* \gamma_k \cdot A_{nijk}   \label{eq:1-gce}
\end{align}
Since $\frac{\lambda_n}{\hat\lambda_\triangle}-1 > 0$ for $n \ge 0$, the eigenvalues of the operator acting on $\gamma_n$ on the left hand side
cannot be less than $\frac{\lambda_n}{\hat\lambda_\triangle}-1$, which is $O(1)$, and a consistent estimate gives rise to
\begin{equation}
\gamma_n = O(\vec\nabla^2\gamma_0,\gamma_0^3).
\end{equation}
for $n\ge 0$. The contribution from $\gamma_n$'s with $n\ge 0$ is beyond the order of the Ginzburg-Landau theory and we have
\begin{align}
  \gamma_0(\vec{x}) = \bigg(\frac{3}{4 \pi}\bigg)^\triangle \ \frac{\braket{\mathcal{O}_\triangle (\vec{x})}}{\sqrt{2}\ \mathrm{U}_\triangle \ T_c^\triangle} \qquad \qquad  \ \text{as  } T \rightarrow T_c
\end{align}
and
\begin{align}
  \Omega(T,\mu;\Psi) = \Omega(T,\mu;0) + \int \mathrm{d}^2\vec x\ \bigg( c \big|\vec\nabla \braket{\mathcal{O}_\triangle (\vec{x})}\big|^2 + a_{_\text{GCE}} |\braket{\mathcal{O}_\triangle (\vec{x})}\big|^2 + \frac{b_{_\text{GCE}}}{2}\big|\braket{\mathcal{O}_\triangle (\vec{x})}\big|^4 \bigg).
\end{align}
The coefficient of gradient term in GL free energy above are:
\begin{align}
  c = \bigg(\frac{3}{4 \pi}\frac{1}{T_c}\bigg)^{2\triangle -3}\frac{Q_{00}}{2\ \mathrm{U}_\triangle^2}.
\end{align}
Numerically, we find that

\begin{equation}
  \begin{aligned}
    \text{for}\ \triangle = 1: \quad    Q_{00} \approx 1.80,\ \   & c \approx 1.23\ T_c \\
    \text{for}\ \triangle = 2: \quad    Q_{00} \approx 4.672,\  & c \approx 0.01\  \frac{1}{T_c} .
  \end{aligned}
\end{equation}

After performing a variation to the order-parameter, we obtain the Ginzburg-Landau equation of an inhomogenous order parameter:
\begin{align}
  -c\ \vec\nabla^2\big|\braket{\mathcal{O}_\triangle(\vec{x})}\big| + a_{_\text{GCE}}(T)\ \big|\braket{\mathcal{O}_\triangle(\vec{x})}\big| + b_{_\text{GCE}}(T)\ \big|\braket{\mathcal{O}_\triangle(\vec{x})}\big|^3 = 0
\end{align}
This formulation can be generalized to the situation with a magnetic field at the boundary by replacing the
ordinary gradient $\vec\nabla$ to the covariant gradient $\vec\nabla-i\vec A$ with $\vec A$ the $U(1)$ vector potential in 2+1 dimensions.

The Ginzburg-Landau formulation in the canonical ensemble with a $\vec x$-independent $\rho$ can be derived similarly and the Helmholtz free energy
with an inhomogenous order parameter reads
\begin{equation}
\begin{aligned}
 F(T,\mu;\Psi) &= \Omega(T,\mu;\Psi) + T\int d ^2\vec x\ \mu\rho \\
 &= F(T,\mu;0) + \int \mathrm{d}^2 \vec x\ \bigg( c \big|\vec\nabla \braket{\mathcal{O}_\triangle(\vec{x})}\big|^2 + a_{_\text{CE}} |\braket{\mathcal{O}_\triangle(\vec{x})}\big|^2 + \frac{b_{_\text{CE}}}{2}\big|\braket{\mathcal{O}_\triangle(\vec{x})}\big|^4 \bigg),
\end{aligned}
\end{equation}
where the Ginzburg-Landau coefficients $a_{_\text{CE}}$ and $b_{_\text{CE}}$ are given by (\ref{eq:coeff-CE}).

\section{Discussions}
\label{sec-7}

In this work, we have developed the Ginzburg-Landau formulation of a holographic superconductor in both the grand canonical
ensemble and the canonical ensemble under the probe limit.
The critical temperature $T_c$ in terms of the chemical potential or the charge density
in our formulation agrees with that reported in the literature \cite{hartnoll_building_2008}. So is the critical exponent of the order
parameter as is expected for a Ginzburg-Landau theory. But our result is not a mere reformulation of existing
knowledge on the subject. Our formula of the constant of proportionality of the critical behavior (\ref{eq:GL_order}), which can be read off from
(\ref{eq:144-CE}), is different from the one in the literature \cite{siopsis_analytic_2010}. The values of this constant in the grand canonical
ensemble and the canonical ensemble are different, reflecting the strong coupling nature of the underlying
superconductivity. In addition, we have also derived the analytical expression of the gradient term.

Let us comment on the difference between our analytical expression for $\kappa$ and that in Ref.\cite{siopsis_analytic_2010}.
We argue that the formulation in \cite{siopsis_analytic_2010} is insufficient to fix this constant of proportionality. Since we made the same
mistake at the early stage of this project, we shall reproduce our experience below. It is convenient to
switch to the dimensionless coordinate $y$ that scales the horizon radius for an arbitrary temperature to one.
At $T=T_c$, $\Psi=0$ and we find the
exact solution to (\ref{eq:eom-phi}),
\begin{equation}
  \Phi=\mu_c(1-y)
\end{equation}
where $\mu_c$ is the chemical potential at the critical temperature. The charge density at $T_c$ reads
$\rho_c=\frac{\mu_c}{z_c}$ with $z_c$ the value of $z_h$ at $T=T_c$. For the temperature $T<T_c$, but
$1-\frac{T}{T_c} \ll 1$, we have $\Psi\approx\gamma_0u_0(y)$ and
\begin{equation}
  \Phi = \mu_c(1-y) + \Delta\Phi
\end{equation}
with the perturbation $\Delta\Phi$ given by the equation
\begin{equation}
  \frac{\mathrm{d}^2\Delta\Phi}{\mathrm{d}y^2} = \frac{2\Psi^2}{y^2 f}\Phi
  \approx 2\mu_c\gamma_0^2\frac{u_0^2(y)}{y^2(1+y+y^2)}.
\end{equation}
Therefore $\Delta\Phi$ is obtained by integrating the right hand side twice. The horizon boundary condition,
$\Delta\Phi=0$ at $y=1$ fixes only one of the two integration constants and we end up with
\begin{equation}
  \Delta\Phi = \alpha\gamma_0^2(1-y) - 2\mu_c\gamma_0^2\int_y^1\mathrm{d}\eta(y-\eta)
  \frac{u_0^2(\eta)}{\eta^2(1+\eta+\eta^2)}.
  \label{eq:pert-phi}
\end{equation}
with the integration constant $\alpha=O(1)$ and
\begin{equation}
  \Phi = (\mu_c + \alpha\gamma_0^2)(1-y) - 2\mu_c\gamma_0^2\int_y^1\mathrm{d}\eta(y-\eta)
  \frac{u_0^2(\eta)}{\eta^2(1+\eta+\eta^2)}.
  \label{eq:sol-phi}
\end{equation}
up to the order $\gamma_0^2=O\left(1-\frac{T}{T_c}\right)$. It follows that the chemical potential
\begin{equation}
  \mu = \mu_c + \alpha\gamma_0^2 + 2\mu_c\gamma_0^2\int_0^1\ \mathrm{d}y\frac{u_0^2(y)}{y(1+y+y^2)}
  \label{eq:mu}
\end{equation}
and the charge density
\begin{equation}
  \rho = \frac{\mu_c+\alpha\gamma_0^2}{z_h} + 2\rho_c\gamma_0^2\int_0^1\ \mathrm{d}y \frac{u_0^2(y)}{y^2(1+y+y^2)}.
  \label{eq:rho-w}
\end{equation}
up to the same order, $O\left(1-\frac{T}{T_c}\right)$. Combining (\ref{eq:mu}) and (\ref{eq:rho-w}) to eliminate $\alpha$, we find that
\begin{equation}
  \rho = \frac{\mu}{z_h} + 2\rho_c\gamma_0^2\int_0^1\mathrm{d}y\frac{(1-y)u_0^2(y)}{y^2(1+y+y^2)}.
\end{equation}
In the canonical ensemble (grand canonical ensemble), the temperature dependence of $\mu$ ($\rho$) away
from $T_c$ is undetermined. Therefore the constant in (\ref{eq:GL_order}) cannot be fixed this way.

Upon substitution of our numerical
solution $u_0(y)$ of the Sturm-Liouville problem, which yields the coefficient $\kappa\simeq 144$ with our formula in the canonical
ensemble, into the formula in \cite{siopsis_analytic_2010}, we find the value 168 for $\kappa$.

The Ginzburg-Landau formulation developed in this work can be readily generalized beyond the probe limit. Let us consider the solution
of the equations of motion for the metric $g_{ab}$ and $\Phi$,
\begin{equation}
  \frac{\delta S_{_\text{HSC}}}{\delta g_{ab}} = 0  \qquad  \frac{\delta S_{_\text{HSC}}}{\delta\Phi} = 0
\end{equation}
with $S_{_\text{HSC}}$ given by (\ref{eq:action-HSC}) as functional of $\Psi$.
For a hairless black hole, $\Psi=0$, $\Phi=\bar\Phi$ and $g_{ab}=\bar g_{ab}$ with $\bar\Phi$ the electrostatic potential
of a charged black hole and $\bar g_{ab}$ the
Reissner-Nordstrom-AdS(RNA) metric. In the presence of a small nonzero $\Psi$, The solutions becomes
\begin{equation}
  g_{ab} = \bar g_{ab} + \Delta g_{ab} \qquad \Phi = \bar \Phi + \Delta\Phi
  \label{perturbation}
\end{equation}
with $\Delta g_{ab} = O(\Psi^2)$ and $\Delta\Phi = O(\Psi^2)$. Substituting (\ref{perturbation}) back to the action (\ref{eq:action-HSC}), we
end up with
\begin{equation}
  S_{_\text{HSC}} = S_{_\text{HSC}}[\bar g,\bar\Phi,\Psi] + O(\Delta g^2)
  + O(\Delta g\Delta\Phi) + O(\Delta\Phi^2)
\end{equation}
It follows from the Einstein equation for $g_{ab}$, (\ref{eq:Einstein_equation}) and the equation (\ref{eq:eom-phi}) for $\Phi$ (in a general metric)
that $\Delta g_{ab} = O(\Psi^2)$ and $\Delta\Phi = O(\Psi^2)$. Therefore beyond the probe limit,
the metric underlying the Sturm-Liouville problem becomes that of a RNA black hole.
The deviations from RNA metric contribute only to the quartic term of the Ginzburg-Landau theory.

The holographic superconductivity discussed here and investigated in the literature all come from the classical limit
of its gravity dual. Following the example of the correspondence between the super Yang-Mills and the
superstring in $AdS_5\times S^5$, it is conceivable that the underlying field theory of the
superconductivity at the boundary is in some large-$N$ limit with $N$ the multiplicity of some internal degrees of freedom.
Then the fluctuation of the long rage order can be ignored and  Ginzburg-Landau formulation developed in this
paper becomes exact. For a realistic superconductor, say finite $N$ case, the long range order at
a nonzero temperature will be completely destroyed by the fluctuation of its phase in a space of dimensions two
or less, in accordance with the well-known Hohenberg theorem \cite{hohenberg_existence_1967}.
Therefore, singularities should
emerge if the holographic correspondence is generalized beyond the large-$N$ limit. Without the knowledge of
the underlying string theory in the bulk and its field theory image at the boundary, one can only suggest some
possibilities. The finite $N$-correction may bring about some stronger metric singularities at the horizon that renders
the Sturm-Liouville problem for the critical temperature non self-adjoint. Or the action (\ref{eq:action})
acquires a non-polynomial potential of the scalar field that is singular at $\Psi=0$.

\acknowledgments
We thank C. Herzog for an email communication. The research of Defu Hou and Hai-cang Ren is supported in part by NSFC under grant Nos. 11375070 ,  11221504, 11135011.
\appendix

\section{The Ginzburg-Landau theory of a BCS superconductor}
\label{sec-6}

In this appendix, we review the Ginzburg-Landau free energy of an ordinary BCS superconductor in the grand canonical and the canonical ensembles
to highlight the difference between the weakly coupled BCS and the strongly coupled HSC.

The mean field Hamiltonian of a BCS superconductor appropriate in the grand canonical ensemble reads
\begin{equation}
\mathcal{H}=V\frac{\Delta^2}{\lambda}+\sum_{\vec{k},\vec{s}}(\frac{k^2}{2m}-\mu) a_{\vec{k},s}^\dagger a_{\vec{k},s}
+\Delta\sum_{\vec{k}}(a_{\vec{k},\uparrow} a_{-\vec{k},\downarrow}+a_{-\vec{k},\downarrow}^\dagger a_{\vec{k},\uparrow}^\dagger),
\end{equation}
where $a_{\vec{k},s}$ and $a_{\vec{k},s}^\dagger$ are annihilation and creation operators of an electron of momentum ${\vec{k}}$ and spin $s(\uparrow,\downarrow)$,
$\Delta$ is the order parameter (energy gap), $\mu$ is the chemical potential, $\lambda>0$ is the pairing coupling,
$m$ is the electron mass and $V$ is the magnetization volume. The grand partition function at the temperature $T$ is
\begin{equation}
{\cal Z}={\rm Tr}\exp\left(-\frac{{\cal H}}{T}\right)
\end{equation}
and the density of the thermodynamic potential is given by
\begin{equation}
\omega=-\frac{T}{V}\ln{\cal Z}.
\end{equation}

The Ginzburg-Landau free energy corresponds to $\omega$ expanded to the quartic power of $\Delta$ near the critical temperature, $T_c$,
We have
\begin{equation}
\omega=\omega_0+a\Delta^2+\frac{1}{2}b\Delta^4
\end{equation}
where $\omega_0$ is the thermodynamic potential density of a free electron gas and the coefficient $a$ and $b$ are given by one-loop thermal diagrams. The weak coupling condition
corresponds to
\begin{equation}
T_c \ll \mu.
\label{weak}
\end{equation}
It is straightforward to show that
\begin{equation}
\omega_0=-\frac{(2m\mu)^{\frac{5}{2}}}{15\pi^2m},
\end{equation}

\begin{equation}
a=\frac{1}{\lambda}-\frac{m^{\frac{3}{2}}\mu^{\frac{1}{2}}}{\sqrt{2}\pi^2}\int_{-\nu_D}^{\nu_D}\frac{\mathrm{d}\xi}{\xi}\tanh\frac{\xi}{2T}
\approx -\frac{m^{\frac{3}{2}}\mu^{\frac{1}{2}}}{\sqrt{2}\pi^2}\left(1-\frac{T}{T_c}\right)
\end{equation}
with $\nu_D$ the Debye frequency ($T_c \ll \nu_D \ll \mu$) and
\begin{equation}
b=\frac{7\zeta(3)}{8\sqrt{2}\pi^4}\frac{m^{\frac{3}{2}}\mu^{\frac{1}{2}}}{T_c^2},
\end{equation}
The charge density
\begin{equation}
\rho=-\left(\frac{\partial\omega}{\partial\mu}\right)_{T,\Delta}=\rho_0-\left(\frac{\partial a}{\partial\mu}\right)_{T,\Delta}\Delta^2+O(\Delta^4)
\label{rho}
\end{equation}
where
\begin{equation}
\rho_0=\frac{(2m\mu)^{\frac{3}{2}}}{3\pi^2}
\end{equation}

\begin{equation}
\left(\frac{\partial a}{\partial\mu}\right)_{T,\Delta}=-\frac{m^{\frac{3}{2}}\mu^{-\frac{1}{2}}}{2\sqrt{2}\pi^2}\left(\ln\frac{\nu_D}{T_c}+{\rm const.}\right)
\end{equation}

Now we are ready to calculate the Helmholtz energy density pertaining to the canonical ensemble via the Legendre transformation
\begin{equation}
f=\omega + \mu\rho
\label{helmholtz}
\end{equation}
It follows from (\ref{rho}) that
\begin{equation}
\mu = \mu_0+\kappa\Delta^2+O(\Delta^4),
\label{inverse}
\end{equation}
where
\begin{equation}
\mu_0=\frac{(3\pi^2\rho)^{\frac{2}{3}}}{2m}
\label{eq:leading-bcs}
\end{equation}
and
\begin{equation}
\kappa=\frac{\left(\frac{\partial a}{\partial\mu}\right)_{T,\Delta}\bigg|_{\mu=\mu_0}}{\frac{\partial\rho_0}{\partial\mu}\bigg|_{\mu=mu_0}}.
\end{equation}
Substituting (\ref{inverse}) into (\ref{helmholtz}), we obtain that
\begin{equation}
f=f_0+a\Delta^2+\frac{1}{2}b\Delta^4+\frac{1}{2}b'\Delta^4+...
\end{equation}
where the quartic term comes from the square of the second term of (\ref{inverse}) and is given by
\begin{equation}
b'=\frac{\left(\frac{\partial a}{\partial\mu}\right)_{T,\Delta}^2\bigg|_{\mu=\mu_0}}{\frac{\partial\rho_0}{\partial\mu}\bigg|_{\mu=\mu_0}}
=\frac{m^{\frac{3}{2}}\mu_0^{-\frac{1}{2}}}{8\sqrt{2}\pi^2}\left(\ln\frac{\omega_D}{T_c}+\mathrm{const.}\right)^2
\end{equation}
It follows from the weak coupling condition (\ref{weak}) that $b' \ll b$ and can be ignored.
The quartic term of a BCS superconductor in the canonical ensemble is therefore simply obtained from that in the grand canonical ensemble
by substitution of the leading order relation (\ref{eq:leading-bcs})and the scaling law of the order parameter $\Delta$ as $T\to T_c$
from below takes the identical form
\begin{equation}
  \Delta = \sqrt{-\frac{a}{b}} = T_c\sqrt{\frac{8\pi^2}{7\zeta(3)}\left(1-\frac{T}{T_c}\right)} = 3.0633\ T_c \sqrt{1 - \frac{T}{T_c}}
\end{equation}
in both ensembles.This, however, is not the case in the strong coupling.

%\bibliographystyle{JHEP}
%\bibliography{MyLibrary}

\begin{thebibliography}{9}

\bibitem{hooft_dimensional_1993}
 G.~t. Hooft,
{\it Dimensional reduction in quantum gravity},[gr-qc/9310026].

\bibitem{susskind_world_1994}
 L.~Susskind,
{\it The world as a hologram},
 J.Math.Phys. {\bf 36} :6377-6396 (1995). [{arXiv}: hep-th/9409089],

\bibitem{maldacena_large_1998}
  J. Maldacena,
{\it The Large N Limit of Superconformal Field Theories and Supergravity},
Adv.  Theor. Math. Phys. {\bf 2}, 231 (1998). [hep-th/9711200]

\bibitem{wittern_anti-de_1998}
  E. Witten,
{\it Anti-de Sitter Space and Holography},
  Adv. Theor. Math. Phys. {\bf 2}, 253 (1998).  [hep-th/9802150]

\bibitem{gubser_gauge_1998}
  S.~S. Gubser, I.~R. Klebanov, and A.~M. Polyakov,
{\it Gauge theory correlators from non-critical string theory},
  Phys. Lett.   B 428 {\bf}: 105-114 (1998). [hep-th/9802109]

\bibitem{klebanov_ads/cft_1999}
 I.~R. Klebanov and E.~Witten,
{\it AdS/CFT correspondence and symmetry breaking},
 Nucl.Phys. B 556 {\bf 1}:89-114 (1999).[hep-th/9905104]

\bibitem{aharony_large<_2000}
 O.~Aharony, S.~S. Gubser, J.~Maldacena, H.~Ooguri, and Y.~Oz,
 {\it  Large $N$ field theories, string theory and gravity},
 Phys. Rept. 323 {\bf 3}, 183-386 (2000).[hep-th/9905111]

% \bibitem{liu_calculating_2006}
%  Liu, Hong, Krishna Rajagopal, and Urs Achim Wiedemann.
% {\it Calculating the jet quenching parameter},
% Phys. Rev. Lett  97.{\bf 18}: 182301 (2006).

\bibitem{policastro_shear_2001}
  G.~Policastro, D.~T. Son, and A.~O. Starinets,
{\it Shear viscosity of strongly coupled $\mathcal{N}= 4$ supersymmetric Yang-Mills plasma},
 Phys. Rev. Lett. 87 {\bf 8}:081601 (2001).[hep-th/0104066]

\bibitem{gubser_breaking_2008}
 S.~S. Gubser,
{\it Breaking an abelian gauge symmetry near a black hole horizon},
 Phys. Rev. D {\bf 78}, 065034 (2008).  [hep-th/0801.2977].

\bibitem{gubser_gravity_2008}
  S.~S. Gubser,
{\it The gravity dual of a p-wave superconductor},
J. High Energy Phys. {\bf 0811}:033 (2008).[hep-th/0805.2960]

\bibitem{hartnoll_building_2008}
  S.~A. Hartnoll, C.~P. Herzog, and G.~T. Horowitz,
 {\it Building an AdS/CFT superconductor},
 Phys. Rev. Lett. {\bf 101}, 031601  (2008).[hep-th/0803.3295]

\bibitem{hartnoll_holographic_2008}
 S.~A. Hartnoll, C.~P. Herzog, and G.~T. Horowitz,
{\it Holographic superconductors},
 J. High Energy Phys. {\bf 0812}:015 (2008).[hep-th/0810.1563].

\bibitem{horowitz_holographic_2008}
 G.~T. Horowitz and M.~M. Roberts,
{\it Holographic superconductors with various condensates},
Phys. Rev. D {\bf 78} 126008 (2008). [hep-th/0810.1077]

\bibitem{herzog}
C. Herzog, P. K. Kovtun and D. T. Son,
{\it Holographic model of superfluidity},Phys. Rev. D {\bf 79}, 066002 (2009).[hep-th/0809.4870]

\bibitem{maeda_characteristic_2008}
 Maeda K and Okamura T,
{\it Characteristic length of an AdS/CFT superconductor}
Phys. Rev. D {\bf 78} 106006 [hep-th/0809.3079]

\bibitem{maeda_universality_2009}
  Maeda K, Makoto N and Okamura T. ,
{\it Universality class of holographic superconductors}
 Phys.Rev.D {\bf 79} 126004 (2009).  [hep-th/0904.1914]

\bibitem{gregory_holographic_2009}
 R.~Gregory, S.~Kanno, and J.~Soda,
{\it Holographic superconductors with higher curvature corrections},
 J. High Energy Phys. {\bf 0910}:010 (2009). [hep-th/0907.3203]

\bibitem{horowitz_zero_2009}
  Horowitz, Gary T., and Matthew M. Roberts. ,
  {\it Zero temperature limit of holographic superconductors}
  J. High Energy Phys. {\bf 0910}:015 (2009).

\bibitem{barclay_gauss-bonnet_2010}
 L.~Barclay, R.~Gregory, S.~Kanno, and P.~Sutcliffe,
{\it Gauss-Bonnet holographic superconductors},
J. High Energy Phys. {\bf 1012}:029 (2010)  [hep-th/1009.1991]

\bibitem{setare_holographic_2010}
 M.~R. Setare, D.~Momeni, and N.~Majd,
{\it Holographic superconductors in a model of non-relativistic gravity},
J. High Energy Phys. {\bf  1105}:118  (2011). [hep-th/1003.0376].

\bibitem{cai_holographic_2010}
 R.-G. Cai, Z.-Y. Nie, and H.-Q. Zhang,
{\it Holographic p-wave superconductors from Gauss-Bonnet gravity},
 Phys. Rev. D {\bf 82}, 066007 (2010). [hep-th/1009.1991].

\bibitem{ge_analytical_2010}
 X.-H. Ge, B.~Wang, S.-F. Wu, and G.-H. Yang,
{\it Analytical study on holographic superconductors in external magnetic field},
J. High Energy Phys. {\bf 1008}:119.  [hep-th/1002.4901].

\bibitem{benini_gauge_2010}
 F.~Benini, C.~P. Herzog, R.~Rahman, and A.~Yarom,
 {\it Gauge gravity duality for d-wave superconductors: prospects and challenges},
J. High Energy Phys. {\bf } {\bf 1011}:135 (2010).

\bibitem{chen_towards_2010}
 J.-W. Chen, Y.-J. Kao, D.~Maity, W.-Y. Wen, and C.-P. Yeh,
{\it Towards a holographic model of d-wave superconductors},
Phys. Rev. D {\bf81}: 106008 (2010).

\bibitem{jing_holographic_2010}
  Jing, J.,  Chen, S. ,
 {\it Holographic superconductors in the Born-Infeld electrodynamics},
 Phys. Lett. B, 686 {\bf 1}, 68-71.(2010).[hep-th/1001.4227]

\bibitem{zeng_d-wave_2010}
 H.-B. Zeng, Z.-Y. Fan, and H.-S. Zong,
 {\it d-wave holographic superconductor vortex lattice and non-abelian holographic superconductor droplet},
 Phys.Rev.D {\bf 82}:126008 (2010). [hep-th/1007.4151].

\bibitem{zeng_analytical_2010}
 H.-B. Zeng, X.~Gao, Y.~Jiang, and H.-S. Zong,
{\it Analytical computation of critical exponents in several holographic superconductors},
 J. High Energy Phys. {\bf 1105}:002,2011. [hep-th/1012.5564].

\bibitem{chen_analytic_2011}
 C.-M. Chen and M.-F. Wu,
{\it An analytic analysis of phase transitions in holographic superconductors},
Prog. Theor. Phys. {\bf 126}, 387-395  (2011).[hep-th/1103.5130].

\bibitem{kanno_note_2011}
 S.~Kanno,
{\it A note on Gauss-Bonnet holographic superconductors},
 Class.Quant.Grav.{\bf 28}:127001, (2011).[hep-th/1103.5022].

\bibitem{jing_holographic_2011}
J.~Jing, Q.~Pan, and S.~Chen,
{\it Holographic superconductors with power-Maxwell field},
J. High Energy Phys. {\bf 1111}:112  (2011).[hep-th/1106.5181]

\bibitem{ge_analytical_2011}
 X.-H. Ge and H.-Q. Leng,
 {\it Analytical calculation on critical magnetic field in holographic superconductors with backreaction},
 Prog. Theor. Phys. {\bf 128} (2012), 1211-1228.[hep-th/1105.4333].

\bibitem{pan_analytical_2012}
 Q.~Pan, J.~Jing, B.~Wang, and S.~Chen,
{\it Analytical study on holographic superconductors with backreactions},
 J. High Energy Phys. {\bf 1206}: 087  (2012).[hep-th/ 1205.3543].

\bibitem{gangopadhyay_analytic_2012-2}
 S.~Gangopadhyay and D.~Roychowdhury,
{\it Analytic study of properties of holographic superconductors in Born-Infeld electrodynamics},
J. High Energy Phys. {\bf 1205}:113 (2012).[hep-th/1201.6520]

%\bibitem{liu_gauss-bonnet_2012}
 %Y.~Liu, Y.~Peng, and B.~Wang,
%{\it Gauss-bonnet holographic superconductors in Born-Infeld electrodynamics with backreactions},


\bibitem{gangopadhyay_analytic_2012}
 S.~Gangopadhyay and D.~Roychowdhury,
{\it Analytic study of Gauss-Bonnet holographic superconductors in Born-Infeld electrodynamics},
 J. High Energy Phys. {\bf 1205}:156  (2012).[hep-th/1204.0673].

\bibitem{momeni_holographic_2013}
 D. Momeni, R. Myrzakulov, and M. Raza,
{\it Holographic superconductors with Weyl corrections via gauge/gravity duality},
 Int.J.Mod.Phys. {\bf A28}: 1350096 (2013).[hep-th/ 1307.8348]

%\bibitem{amado_holographic_2013}
 %I.~Amado, D.~Arean, A.~Jimenez-Alba, L.~Melgar, and I.~S. Landea,
 %{\it  Holographic s+p superconductors},  [hep-th/1309.5086].

\bibitem{yao_analytical_2013}
 W.~Yao and J.~Jing,
{\it Analytical study on holographic superconductors for  Born-Infeld electrodynamics in Gauss-Bonnet gravity with backreactions},
 J. High Energy Phys. {\bf 1305} (2013) 101.[hep-th/1306.0064]

\bibitem{li_gauss-bonnet_2013}
  Li, Yong-Zhuang, Shao-Feng Wu, and Guo-Hong Yang ,
{\it Gauss-Bonnet correction to holographic  thermalization: two-point functions, circular Wilson loops and entanglement  entropy},
Phys. Rev. D {\bf88.8} (2013): 086006. [hep-th/ 1309.3764].

\bibitem{siopsis_analytic_2010}
 G.~Siopsis and J.~Therrien,
 {\it Analytic calculation of properties of   holographic superconductors},
 J. High Energy Phys. .{\bf 1005}: 013 (2010).[hep-th/1003.4275]

\bibitem{herzog_analytic_2010}
  Herzog, C. P. {\it Analytic holographic superconductor}, Phys. Rev. D {\bf 81}.12 (2010): 126009.[hep-th/1003.3278]

\bibitem{henneaux2004asymptotically}
 Henneaux, Marc, et al.
{\it Asymptotically anti–de Sitter spacetimes and scalar fields with a logarithmic branch.} Phys. Rev. D {\bf 70}.4 (2004): 044034.

\bibitem{henneaux2007asymptotic}
Henneaux, Marc, et al.
{\it Asymptotic behavior and Hamiltonian analysis of anti-de Sitter gravity coupled to scalar fields.}
 Annals of Phys. {\bf 322}.4 (2007): 824-848.

\bibitem{witten2001multi}
 Witten, Edward.
{\it Multi-trace operators, boundary conditions, and AdS/CFT correspondence.}
 arXiv preprint hep-th/0112258 (2001).

\bibitem{hartnoll_lectures_2009}
  Hartnoll, Sean A.,
  {\it Lectures on holographic methods for condensed matter physics},
  Class.  Quant. Grav. .{\bf 26}: 224002  (2009).

\bibitem{horowitz_introduction_2010}
 G.~T. Horowitz,
 {\it Introduction to holographic superconductors},
 [hep-th/ 1002.1722].

\bibitem{brown_thermodynamic_1990}
 Brown, J. David, et al. ,
{\it Thermodynamic ensembles and gravitation},
Class.  Quant. Grav. 7.{\bf 8} (1990): 1433.

\bibitem{braden_charged_1990}
Braden, H. W., Brown, J. D., Whiting, B. F. and  York Jr, J. W. ,
{\it Charged black hole in a grand canonical ensemble},
Phys. Rev. D, 42({\bf 10}), 3376 (1990).

%\bibitem{brown_path_1994}
 %J.~Brown and J.~W. York,
 %{\it The path integral formulation of gravitational thermodynamics} ,
%[ gr-qc/9405024].

\bibitem{hohenberg_existence_1967}
 P.~C. Hohenberg,
{\it Existence of long-range order in one and two dimensions},
Phys. Rev.  158. {\bf 2} : 383-386 (1967).


\end{thebibliography}

\end{document}